# Thermalization of the system "$^3$He-aerogel" at 1.5 K

E.M. Alakshin[1], R.R. Gazizulin[1], A.V. Klochkov[1], V.V. Kuzmin[1], M.S. Tagirov[1], D.A. Tayurskii[1], N. Mulders[2]

[1] Institute of Physics, Kazan (Volga region) Federal University, Kazan, Russia
[2] Physics and Astronomy Department, University of Delaware, Newark, USA

At the present work we propose a new method for studying the processes of thermodynamic equilibrium setting in the adsorbed $^3$He film in a porous media. By this method we have studied the thermalization of the adsorbed $^3$He on the silica aerogel surface at the temperature 1.5 K. The process of the thermodynamic equilibrium setting was controlled by measuring of the pressure in the experimental cell, the amplitude of the NMR signal and the time of the nuclear spin-spin and spin-lattice relaxation times of an adsorbed $^3$He. The thermodynamic equilibrium setting in the system "adsorbed helium-3 – aerogel" has the characteristic time 26 min.

*submitted to arXiv 10$^{th}$ of December 2010*

At present time the influence of an aerogel on properties of superfluid $^3$He is the matter of interest [1,2]. Unique structure of an aerogel leads to anomaly small thermal conductance and the question about real temperature of aerogel still has not been solved. In this paper we report experimental data on this subject.

The aerogel sample (95%) was cut out in the shape of cylinder (d=5mm, h=12mm) and was sealed leak tight in the glass tube (pyrex) to the gas handling system. The temperature of NMR cell has been controlled by $^4$He vapor pumping and in all presented experiments was 1.5 K.

The longitudinal magnetization relaxation time $T_1$ of $^3$He was measured by the saturation recovery method using FID signal. The spin-spin relaxation time $T_2$ was measured by Hahn method. The hand made pulse NMR spectrometer has been used (frequency range 3 – 50 MHz). The pulse NMR spectrometer is equipped by resistive electrical magnet with a magnetic field strength up to 1T. The necessary amount of $^3$He for full coverage of an aerogel surface was adjusted like in [3,4]. On all experimental data (Fig.1-4) the time scale is the time after beginning of $^3$He condensation procedure.

The process of $^3$He condensation (adsorption) consists of two exponential processes (Fig.1). We can suggest, the first one is the process of preliminary adsorption of $^3$He molecules at an aerogel surface (characteristic time τ=2min). The next process is the redistribution of $^3$He molecules in adsorbed layer (characteristic time τ=15min) and further pressure decreasing because of an additional adsorption.

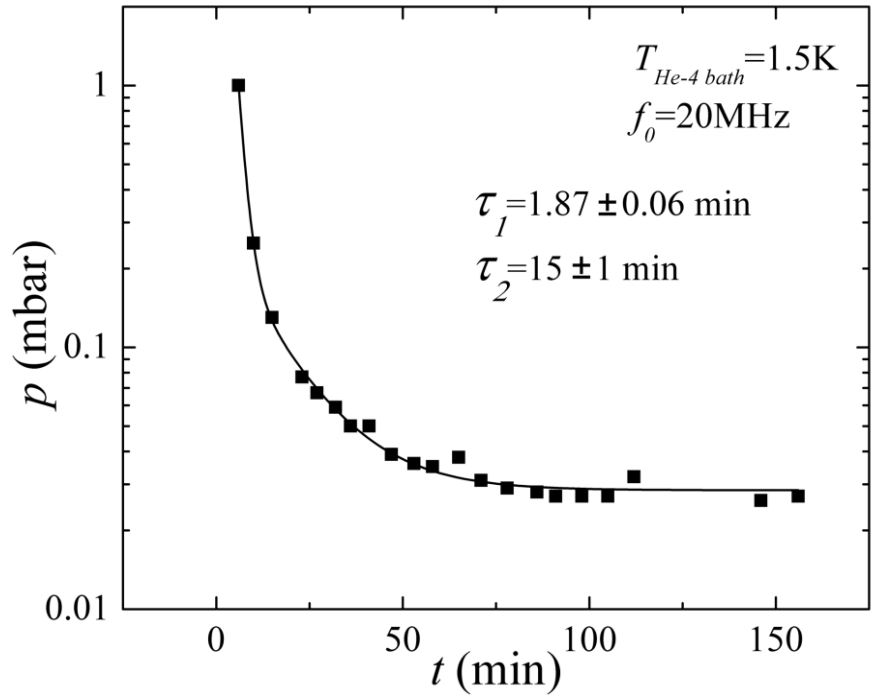

Fig.1 The process of $^3$He condensation in aerogel.

On the Fig.2 the evolution of $^3$He spin echo amplitude in the same time frame as on Fig.1 is presented. The characteristic time of this process is about 26 min.

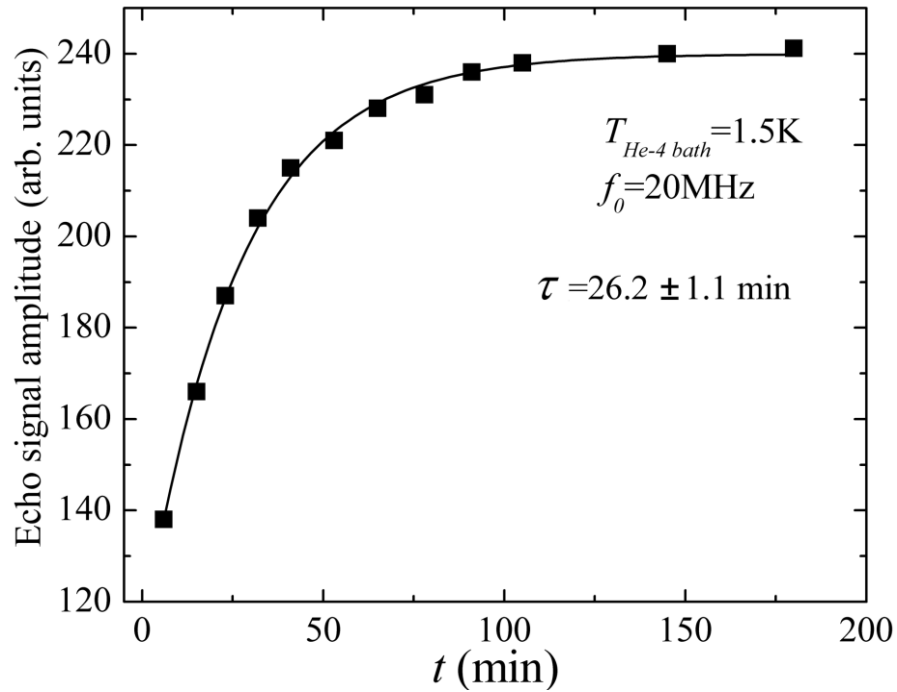

Fig.2 The evolution of $^3$He spin echo amplitude during thermalization process.

On the Fig.3 and Fig.4 the evolution of transverse relaxation time $T_2$ and longitudinal relaxation time $T_1$ are presented. About the same characteristic times are observed.

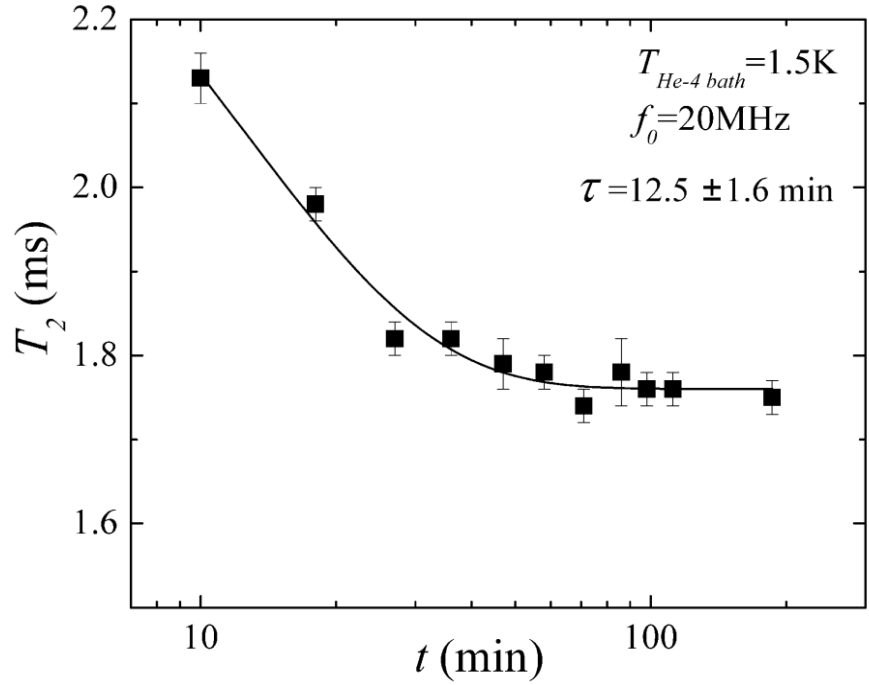

Fig.3 The evolution of transverse relaxation time $T_2$ of $^3$He during thermalization process

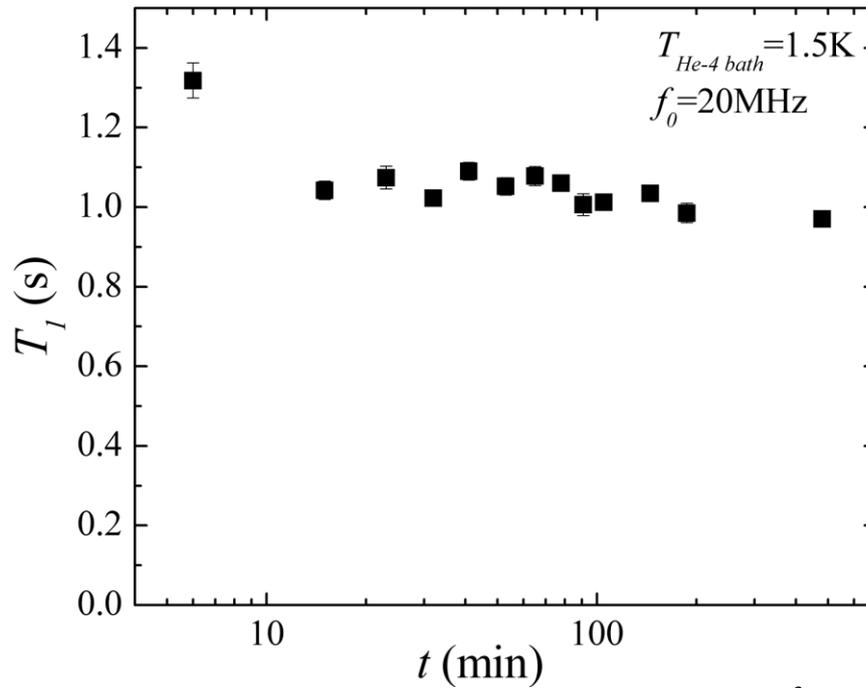

Fig.4 The evolution of longitudinal relaxation time $T_1$ of $^3$He during thermalization process

All presented experimental data show that thermalization processes in the system "$^3$He-aerogel" takes time less than 25 min. We didn't observe any longer processes in thermalization. Taking into account the fact, that NMR cell with an aerogel was cooled down to 1.5K within 2 hours prior of $^3$He condensation we can conclude, that the temperature of "$^3$He-aerogel" system became 1.5K.

Also, experimental data shows that thermalization process occurs in three steps. Fastest process is the process of preliminary adsorption of $^3$He molecules at an aerogel surface (characteristic time is about 2 min.). The next step is the redistribution of $^3$He molecules in adsorbed layer, which accompanies by setting to equilibrium values for transverse relaxation time $T_2$ and longitudinal relaxation time $T_1$ $^3$He (characteristic time of these processes is about 15 min.). The last step is the process of achieving the global thermal equilibrium between "adsorbed $^3$He-aerogel" system and $^4$He bath (characteristic time 26 min.).

At the present work only particular case of the system "$^3$He-aerogel" was considered, namely $^3$He adsorbed layer and its thermalization. If free atoms of $^3$He would be introduced into the system (gas or liquid phases) the thermalization processes would be accelerated by fast mass diffusion and energy exchange between adsorbed layer and NMR cell walls. For example, an addition of small gas portion of $^3$He (5 mbar pressure) above completed adsorbed layer will accelerate thermalization process up to 5 min.

As a result we propose a new method for studying the processes of thermodynamic equilibrium setting in the adsorbed $^3$He film in a porous media. By this method we have studied the thermalization of the adsorbed $^3$He on the silica aerogel surface at the temperature 1.5 K.

This work is partially supported by Russian Fund for Basic Research (grant N 09-02-01253).

1. Porto J and Parpia, *J Phys. Rev. Lett.* **74**, 4667 (1995)
2. Sprague D, Haard T, Kycia J et al., *Phys. Rev. Lett.* **75**, 661 (1995)
3. A. V. Klochkov, V. V. Kuzmin, K. R. Safiullin et al., *JETP Letters*, **88**, 823 (2008)
4. A. Klochkov, V. Kuzmin, K. Safiullin et al., *Journal of Physics: CS*, **150**, 032043 (2009)